\documentclass[showpacs,floatfix,superscriptaddress,showpacs,preprint,amssymb,amsfonts,prb,aps]{revtex4}
\usepackage{longtable,graphicx,epsfig,dcolumn}

\begin{document}
\bibliographystyle{revtex}
\title{Surface Induced Order in Liquid Metals and Binary Alloys}

\author{Elaine~DiMasi}
\affiliation{Department of Physics, Brookhaven National Laboratory,
Upton NY 11973-5000}

\author{Holger~Tostmann}
\affiliation{Division of Applied Sciences and Department of Physics,
Harvard University, Cambridge MA 02138}

\author{Oleg~G.~Shpyrko}
\affiliation{Division of Applied Sciences and Department of Physics,
Harvard University, Cambridge MA 02138}

\author{Peter~S.~Pershan}
\affiliation{Division of Applied Sciences and Department of Physics,
Harvard University, Cambridge MA 02138}

\author{Benjamin~M.~Ocko}
\affiliation{Department of Physics, Brookhaven National Laboratory,
Upton NY 11973-5000}

\author{Moshe~Deutsch}
\affiliation{Department of Physics, Bar-Ilan University, Ramat-Gan
52100, Israel}

\begin{abstract}

Surface x-ray scattering measurements from several pure liquid
metals (Hg, Ga and In) and from three alloys (Ga-Bi, Bi-In, and
K-Na) with different heteroatomic chemical interactions in the
bulk phase are reviewed. Surface induced layering is found for
each elemental liquid metal. The surface structure of the K-Na
alloy resembles that of an elemental liquid metal. Surface
segregation and a wetting film are found for Ga-Bi. Bi-In displays
pair formation at the surface.

\end{abstract}

\pacs{61.25.Mv,~68.10.--m,~61.10.--i}

\maketitle

\section{Liquid Metals and Surface Induced Order}

Liquid metals (LM) are comprised of charged ion cores whose
Coulomb interactions are screened by a conduction electron sea. At
the liquid-vapor interface, this screened Coulomb potential gives
way to the weaker van der Waals interactions that prevail in the
vapor. Since the potential changes so substantially across the
interface, the potential gradient is high, producing a force that
acts on the ions at the liquid surface as though they were packed
against a hard wall. Analytic calculations and molecular dynamics
simulations predict that atoms at the LM surface are stratified in
layers parallel to the
        interface \cite{rice96}.
By contrast, a monotonic density profile is
predicted for the vapor interface of a nonmetallic liquid.

\begin{figure}[tbp]
\centering
\includegraphics[angle=0,width=0.6\columnwidth]{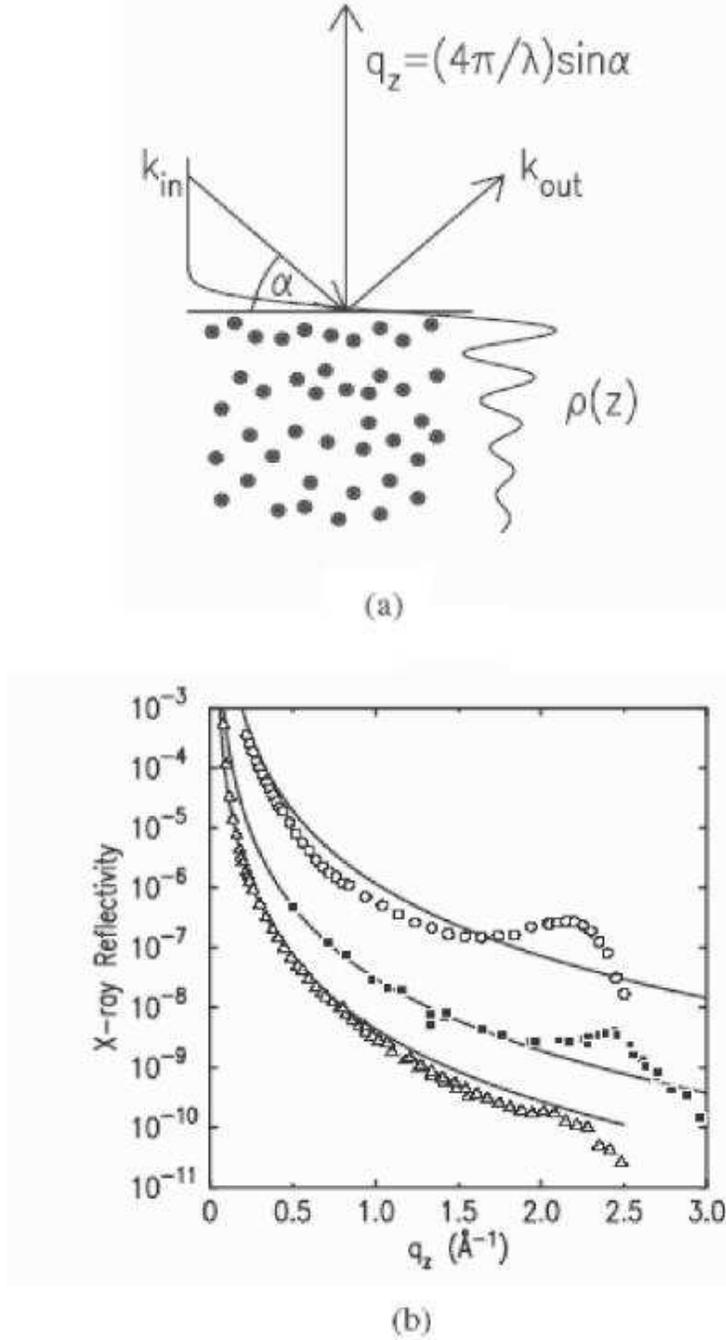}
 \caption{ (a) X-ray reflectivity geometry for the
liquid metal, with layering of ions producing an oscillatory density
profile $\rho(z)$. (b) X-ray reflectivity for liquid Hg
($-35^\circ$C, $\circ$), Ga ($+25^\circ$C) and In ($+170^\circ$C,
$\triangle$). Solid lines: calculated Fresnel reflectivity from a
flat surface. Data for Ga and In are shifted for
clarity.}\label{jop_fig1}
\end{figure}

Observation of surface layering in LM requires an experimental
technique sensitive to the surface-normal density profile that can
resolve length scales of 2--3 \AA. Specular X-ray reflectivity
provides the most direct probe of the surface normal structure.
X-rays incident on the liquid surface at an angle $\alpha$ are
scattered at the same angle within the reflection plane defined by
the incident beam and the surface normal (Fig.~\ref{jop_fig1}(a)).
The reflected intensity is directly related to the surface normal
density profile $\tilde{\rho}(z)$:
\begin{equation}
R ( q_z) \propto \left|  q_z^{-2} ( \partial \tilde{  \rho } (z )
/\partial z) \exp (iq_z z) dz \right|^2 .
\end{equation}
Since $ \partial \tilde{  \rho }(z )/ \partial z$ is nonzero only
near the surface, x-ray reflectivity is sensitive to the
surface-normal structure and not to the structure of the bulk
liquid. For example, surface layering with a spacing $d$  produces
a quasi-Bragg peak in the reflectivity,  centered at the
surface-normal momentum transfer $q_z = (4 \pi /\lambda)\sin
\alpha \approx 2 \pi /d$
    \cite{bosio,magnu95,regan95}.

Grazing incidence diffraction (GID) is sensitive to the in-plane
structure of the surface. The in-plane momentum transfer $q_{\|}$
is probed by varying the azimuthal angle $2\theta$ at fixed
$\alpha$. This geometry is surface sensitive when the incident
angle $\alpha$ is kept below the critical angle for total external
reflection, $\alpha_{c}$, thereby limiting the x-ray penetration
depth
    \cite{eisenberger}.

For these structural studies it is essential to maintain a liquid
metal surface that is flat and clean on an atomic scale. The
sample is contained either in an ultra high vacuum (UHV)
environment, or under a reducing atmosphere such as dry hydrogen
gas,  to prevent oxidation. For low vapor pressure, UHV-compatible
metals such as Ga, Bi, and In, argon ion sputtering is possible,
and this is the most reliable way to produce an atomically clean
surface\,\cite{regan95}.

Surface layering in elemental LM
was first experimentally confirmed by synchrotron x-ray
reflectivity measurements of liquid
        Hg \cite{magnu95}
and
        Ga \cite{regan95}.
Experiments on
        In \cite{tostmann99}
and a number of
        alloys \cite{lei96,lei97,regan97,tostmann98}
followed.
        Fig.~\ref{jop_fig1}(b) shows
experimental reflectivities for  three low melting point elemental
LM. The principle deviation from the Fresnel reflectivity calculated
for a perfectly flat metal surface (solid lines) is in the broad
quasi-Bragg peak centered near $q_z = 2.2$~\AA$^{-1}$. These
reflectivity profiles can be well described by layered density
profiles decaying over several layers, shown schematically in
Fig.~\ref{jop_fig1}(a).

\section{Surface Structure of Binary Liquid Alloys}

In binary alloys properties such as atomic size, surface tension,
and electronic structure can be varied and should affect the details
of the  surface structure, thus  allowing a more systematic
understanding of  surface layering. Also, since binary alloys form
various ordered phases in the bulk, another interesting question
arises: How does the alloy's bulk phase behavior manifest itself at
the surface, where the electronic structure, atomic coordination and
local composition are different? This question has motivated a
number of studies on alloys, which have found that in general,
surface layering competes with the formation of more complicated
surface phases. For example, in miscible alloys the Gibbs adsorption
rule predicts that the species having the lower surface energy will
segregate at the surface. Observations on
    Ga-In \cite{regan97},
    Ga-Sn \cite{lei97}
and Ga-Bi at low Bi concentrations
    \cite{lei96,tostmann98}
have found that surface segregation coexists with surface
layering.  In these alloys the first surface layer is almost
entirely composed of the lower surface tension component (In, Sn
or Bi). By the second or third atomic layer, the bulk composition
has been reached. In the following sections, we describe recent
x-ray results from alloy surfaces which demonstrate a range of
different surface induced structural effects.

\subsection{K-Na}

Alkali metals have a simple electronic structure which can be
described by ideal Fermi surfaces, and are soluble in each other
with only a weak tendency towards phase formation. Since alkali
metals have a very low surface tension, surface fluctuations are
enhanced. These properties are expected to make the alkali metals'
surface structures different from those of the main group metals
studied so far. Ideally alkali metals would be investigated under
UHV conditions due to their high reactivity. However, at the
melting point their high vapor pressures precludes this. By
contrast, the melting point of the eutectic $K_{80}Na_{20}$ alloy
is sufficiently low to allow UHV conditions. Due to the almost
identical electron densities of the two components, when probed by
x-rays this alloy exhibits the structure of a homogeneous liquid
metal. Here we present preliminary results for the eutectic $
K_{80}Na_{20}$ alloy.

\begin{figure}[tbp]
\centering
\includegraphics[angle=0,width=0.6\columnwidth]{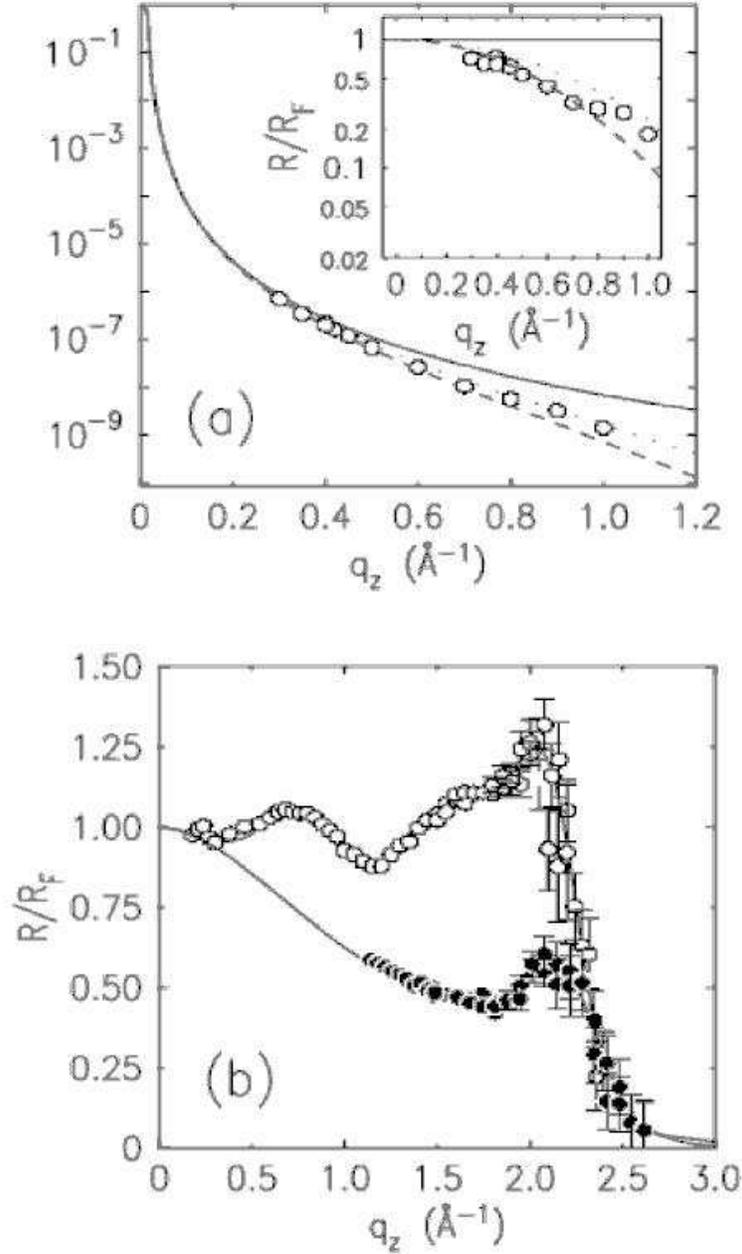}
 \caption{(a) X-ray reflectivity from a $
K_{80}Na_{20}$ alloy measured by integrating over a large range of
$\alpha$ at fixed $\alpha +\beta$. The normalized reflectivity is
shown in the inset.  The dotted lines show a capillary wave
roughness with no layering with $\sigma = 1.2$ and 1.5 \AA,
respectively. (b) Normalized x-ray  reflectivity of liquid In
(+170$^\circ$C, closed triangles) and In-22at \%Bi (80 $^\circ$C,
open triangles).}\label{jop_fig2}
\end{figure}

Fig.~\ref{jop_fig1}(a) shows the x-ray reflectivity from
$K_{80}Na_{20}$ along with the predicted reflectivity assuming
capillary wave roughness (Gaussian form) of 1.2 and
1.5~\AA\cite{oops}. At all $q_z$ the reflectivity is bounded by
these two curves; at lower $q_z$ it is better described by the
1.5~\AA\ roughness.  On length scales $\stackrel{>}{\scriptsize
\sim} 6$~\AA\ no obvious structural feature is found beyond the
predicted capillary wave roughness. The low surface tension
($\approx 120~dyn/cm$) and the subsequently high roughness, appears
to preclude measurements to $q_z$ large enough to directly observe a
surface layering peak. This is in contrast to well-defined surface
layering peaks observed for Ga, Hg or In
   (see Fig.~\ref{jop_fig1}(b)) \cite{EDHT}.

\subsection{Bi-In}

For systems having significant attractive interactions between
unlike atoms, the surface structure is more complex. This is
especially true of alloys such as Bi-In which form well-ordered
intermetallic phases in the bulk solid. In Fig.~\ref{jop_fig2}(b) we
show the normalized reflectivity for the eutectic composition
Bi$_{22}$In$_{78}$, measured at 80$^\circ$C ($\triangle$) along with
the normalized reflectivity for liquid In at $+170^\circ$C. The
alloy exhibits a well defined layering peak centered at
2.0~\AA$^{-1}$ which resembles the layering peak found for pure In
({\bf smudges}). In addition, the reflectivity displays a modulation
with a period of about 0.9~\AA$^{-1}$. This oscillation indicates
that ordering over a short region at the surface occurs with a
length scale nearly twice that of the longer-range layering.  This
suggests the presence of Bi-In pairs at the surface.  A full report
on the phase behavior of three different In-Bi alloys will be given
elsewhere \cite{EDHT}.

\subsection{Ga-Bi}

The Ga-Bi system  is an example of an alloy with repulsive
heteroatomic interactions leading to a bulk miscibility gap. Below
the monotectic temperature, $T_{mono} = 222^\circ$C, a Ga-rich
liquid coexists with a solid Bi phase
    \cite{nattland96}.
However, due to its lower surface energy a Bi monolayer is
expected to segregate at the surface of the Ga-rich liquid. Above
$T_{mono}$, Ga-Bi exhibits a thick wetting film, as predicted for
all binary mixtures with critical
    demixing \cite{cahn77}.
This transition occurs at a characteristic wetting temperature
$T_w$ below the critical temperature
    $T_{crit}$ \cite{nattland96}.
Above $T_w$, a macroscopically thick Bi-rich phase is expected to
completely wet the less dense Ga-rich phase in defiance of
gravity. The Bi concentration in the Ga-rich phase increases with
increasing temperature as long as the Ga-rich liquid coexists with
the solid Bi phase.

\begin{figure}[tbp]
\centering
\includegraphics[angle=0,width=0.6\columnwidth]{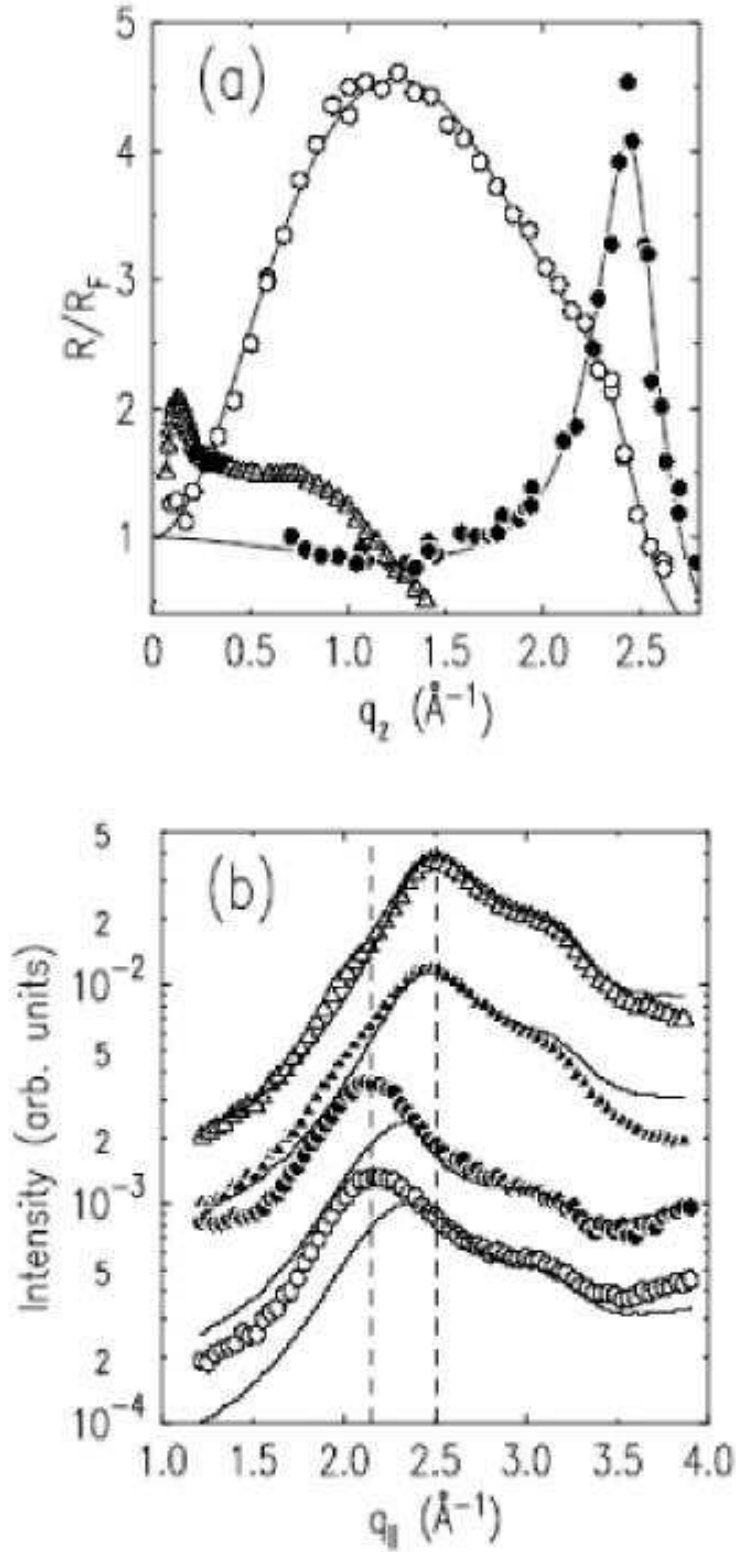}
 \caption{(a) Normalized x-ray reflectivity of
liquid Ga ($+25^\circ$C, $\circ$), and on the Ga-Bi two-phase
coexistence curve at $35^\circ$C ($\bullet$) and $228^\circ$C.
(b)Grazing incidence diffraction from Ga-Bi at $150^\circ$C (filled
triangles), $205^\circ$C ($\triangle$), $228^\circ$C, and
$255^\circ$C ($\Box$) at $\alpha=0.08^\circ$. The solid line shows
corresponding profiles for $\alpha=0.30^\circ$ where the bulk is
predominately sampled. The data was acquired using Soller slits,
0.05 $\AA^{-1}$ FWHM. Still, it was not possible to reliably
subtract the background.}\label{jop_fig3}
\end{figure}

The normalized x-ray reflectivity spectra, $R/R_F$, for Ga-Bi at
$35^\circ$C and $228^\circ$C  are shown in
    Fig.~\ref{jop_fig3}(a)
versus $q_z$, along with the profile for pure Ga at room
temperature. At $35^\circ$C the normalized reflectivity has a
broad maximum at $q_z \approx 1$~\AA$^{-1}$. As suggested by
    Lei et al.\cite{lei96}, this is consistent
with a density profile with a thin, high density monolayer of Bi.

We have fit the reflectivity profiles to simple density profiles
using Eq. (1). The fitted reflectivities are shown in
    Fig.~\ref{jop_fig3}(a) (solid lines).
At $+35^\circ$C the local density profile exhibits a top-layer
density which is about 1.5 times higher than the Ga bulk liquid
density. The $3.4 \pm 0.2 $~\AA\ layer spacing between the surface
and the adjacent Ga layer obtained from the fits is much larger
than the $2.5 \pm 0.1$~\AA\ layer spacing obtained in liquid
gallium. The data show that the surface layer has a higher density
than in the underlying Ga-rich subphase, confirming the surface
segregation of a Bi monolayer.

The behavior of the same alloy at
    $228^\circ$C is markedly
different: a sharp peak in $R(q_z)$ has emerged, centered around
$0.13$~\AA$^{-1}$
    (Fig.~\ref{jop_fig3}(a)).
The peak at small $q_z$ indicates the presence of a thick surface
layer with a density greater than that of the bulk subphase. The
absence of additional oscillations following the sharp peak
suggests that the boundary between the two regions must either be
diffuse or rough. The persistence of the broad maximum at $q_z
\approx 0.75$\AA $^{-1}$ indicates that Bi monolayer segregation
coexists with the newly formed wetting film. Fits to a simple
two-box model yield a film thickness of 30~\AA\, consistent with
ellipsometry results
    \cite{nattland96},
and a surface density consistent with the high density liquid
phase of the bulk alloy. The temperature dependent reflectivity
will be reported elsewhere
    \cite{EDHT}.

In Fig.~\ref{jop_fig3}(b) GID data are shown from the same Ga-Bi
alloy in the temperature range from 150 to 255$^\circ$C. Data at
35$^\circ$C was previously reported
    \cite{tostmann98}.
In the liquid Ga-rich phase the Bi concentration ranges from 3.3
at\% to 17.8at\%. At each temperature data was taken above and
below $\alpha_{crit}= 0.14^\circ $ at $\alpha =0.08^\circ$
(symbols) and at $\alpha=0.30^\circ$(lines). At $\alpha
=0.08^\circ$ the x-ray penetration depth equals $28 \AA$ or about
10 atomic layers.

The solid lines in Fig.~\ref{jop_fig3}(b) at 150 $^\circ$C show
the bulk liquid scattering which is predominately from pure Ga
since the Bi concentration is low. The broad peak at $q_{\|} =
2.5$~\AA$^{-1}$ and the shoulder on the high-angle side of the
peak are in agreement with the bulk liquid Ga structure factor
    \cite{narten72}.

There is no evidence for a peak or shoulder at the position
corresponding to the first peak of the Bi liquid structure factor
at $q_{\|} \approx 2.2$~\AA$^{-1}$.  This is expected since the
surface regime is so much smaller than the bulk volume sampled.
For $\alpha = 0.08^\circ < \alpha_c$, the x-rays penetrate to a
depth of only about 30~\AA . Here a shoulder appears on the
low-$q$ side of the gallium liquid peak, due to enhanced
sensitivity to the Bi surface monolayer. Between $150^\circ$C and
$205^\circ$C ($\bullet$) there is little change in the GID data,
except a slight increase in the shoulder associated with the Bi
monolayer.

Above $T_{mono}$ there is a dramatic change in the GID profiles.
In Fig.~\ref{jop_fig3}(b), GID data is shown at $228^\circ$C and
$255^\circ$C. In both cases, for $\alpha > \alpha_c$ the peak has
shifted to $q_{\|} \approx 2.3$~\AA$^{-1}$ from the 2.5~\AA$^{-1}$
peak position found at lower temperatures.  This results from the
much higher Bi concentration in the bulk at the higher
temperatures and the larger atomic size of Bi.  Even more dramatic
is the shift in the peak position for $\alpha < \alpha_{crit}$
where the peak is at 2.15~\AA$^{-1}$. Thus, the surface region
contains considerably more Bi than the underlying bulk alloy. This
finding is consistent with the wetting layer observed in the x-ray
reflectivity measurements.

\section{Acknowledgments}

This work is supported by the U.S.~DOE Grant No.
DE-FG02-88-ER45379, the National Science Foundation Grant No.
DMR-94-00396 and the U.S.--Israel Binational Science Foundation,
Jerusalem. Brookhaven National Laboratory is supported by U.S. DOE
Contract No. DE-AC02-98CH10886. HT acknowledges support from the
Deutsche Forschungsgemeinschaft.


\begin{thebibliography}{99}
\bibitem{oops} Because the sample surface
was curved, these measurements required an integration over a
range of $\alpha$ at each $q_z$ point.
\end{thebibliography}
\end{document}